\documentclass[twocolumn,showpacs,amsmath,amssymb,floatfix]{revtex4}
\usepackage{graphicx}
\usepackage{dcolumn}
\usepackage{bm}
\begin{document}
\title{Truncation of power law behavior
in ``scale-free'' network models due to 
information filtering}
\author{Stefano Mossa$^{1,2}$, Marc Barth\'el\'emy$^3$, H. Eugene
Stanley$^1$, and Lu\'{\i}s A. Nunes Amaral$^1$}
\affiliation{ $^1$ Center for Polymer Studies and Department of Physics, Boston
University, Boston, Massachusetts 02215 \\ 
$^2$ Dipartimento di Fisica, INFM UdR, and INFM Center for
Statistical Mechanics and Complexity, 
Universit\`{a} di Roma ``La Sapienza'', Piazzale Aldo
Moro 2, I-00185, Roma, Italy\\
$^3$ CEA-Service de Physique de la
Mati\`ere Condens\'ee, BP12, 91680 Bruy\`eres-le-Ch\^atel, France}
\date{\today}
\begin{abstract}
We formulate a general model for the growth of scale-free networks under
filtering information conditions---that is, when the nodes can process
information about only a subset of the existing nodes in the
network.
We find that the distribution of the number of incoming
links to a node follows a universal scaling form, i.e., that it
decays as a power law with an exponential
truncation controlled not only by the system size but also by a feature not
previously considered, the subset of the network ``accessible'' to the
node.  We test our model with empirical data for the World Wide Web and
find agreement.
\end{abstract}
\maketitle
%
%
%

There is a great deal of current interest in understanding the
structure and growth mechanisms of global networks
\cite{Strogatz01,BarabasiRMP,MendesRev} such as the world-wide-web
(WWW) \cite{Albert99,WWW} and the Internet \cite{Internet}. Network structure
is critical in many contexts such as Internet attacks
\cite{BarabasiRMP}, spread of e-mail virus \cite{Cohen94} or dynamics
of human epidemics \cite{MayBook}. In all these problems, the nodes
with the largest number of links play an important role on the
dynamics of the system. It is therefore important to know the global
structure of the network as well as its precise distribution of number
of links.

Recent empirical studies report that both the Internet and the WWW
have scale-free properties, that is, the number of incoming links and
the number of outgoing links at a given node have distributions that
decay with power law tails \cite{Albert99,WWW,Internet}.  It has been proposed
\cite{Barabasi99} that the scale-free structure of the Internet and
the WWW may be explained by a mechanism referred to as ``preferential
attachment'' \cite{Simon} in which new nodes link to existing nodes
with a probability proportional to the number of existing links to
these nodes. Here we focus on the {\it stochastic\/} character of the preferential
attachment mechanism, which we understand in the following way: New nodes
want to connect to the existing nodes with the largest number of
links---i.e., with the largest degree---because of the advantages
offered by being linked to a well-connected node.  For a
large network it is not plausible that a new node will know the degrees
of all existing nodes, so a new node must make a decision on which node to
connect with based on what information it has about the state of
the network.  The preferential attachment mechanism then comes into
play as nodes with larger degree are more likely to become known.

This picture has one underlying and unstated assumption, that
the new nodes will process (i.e., gather, store, retrieve and analyze)
information concerning the state of the entire network.  For very
large networks, such as the WWW or the scientific literature, this
would correspond to the unrealistic situation in which new nodes can
process an extremely large amount of information---i.e., have
unlimited information-processing capabilities.  Indeed, it is likely
that nodes have limited information-processing capabilities and so
must filter incoming information according to their particular
``interests''. Thus, new nodes of a large growing network
will only process information concerning a subset of existing nodes,
since there is a cost associated with processing information. The new
nodes will then make decisions on with whom to link, based on filtered
information. From the standpoint proposed here, most models studied in
the literature work under the unrealistic assumption of {\it
unfiltered\/} information---i.e., a new node processes information about
all the existing nodes in the network.

\begin{figure}[t]
\centering
\includegraphics[width=.4\textwidth]{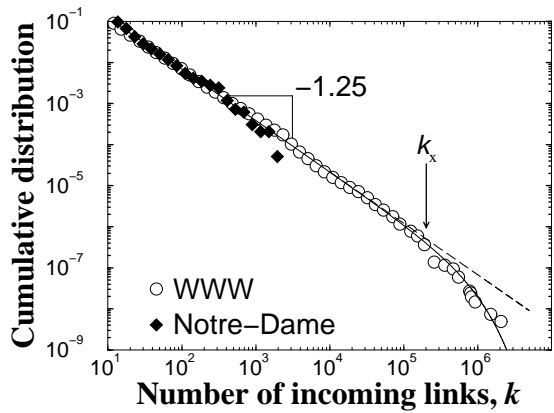}
\vspace{-0.4cm}
\caption{\label{fig1} Distribution of number of incoming links for the WWW.  
Cumulative in-degree distribution from two databases, 
the entire Web \protect\cite{Barabasi99}, and the University
of Notre Dame domain \protect\cite{Albert99}. 
We also plot a power law function with exponent
$\gamma_{in} =1.25$ (dashed line) 
and a Yule function \protect\cite{Simon} of the
form $k^{-\gamma_{in}}\exp(-\alpha k)$ (solid line). 
A cut-off degree $k_\times\simeq$ 200,000 is visible in the data.}
%
\end{figure}

Here we consider for the first time the effect 
on network growth of filtering information due
to limited information-processing capabilities.  
First, we calculate the in-degree distributions of web-pages using two
databases.  The first database, which comprises $\approx 2 \times 10^8$ 
pages \cite{Barabasi99}, surveys a very significant fraction of the entire WWW,
while the second, which comprises $\approx 3\times 10^5$ pages, lists the
University of Notre Dame domain \cite{Albert99}---i.e. the set of URLs
containing the string {\sffamily ``nd.edu''}. For the first database, we
calculate the cumulative in-degree distributions
$P(k) = \sum_{k'>k} p(k')$
where $p(k)$ is the probability distribution.
We confirm that the in-degree distribution decays as
a power law \cite{Barabasi99} of the form 
\begin{equation}
P(k)\sim k ^{-\gamma_{in}}
\label{powerlawe}
\end{equation}
with an exponent $\gamma_{in}=1.25\pm 0.05$ (Fig.~\ref{fig1}). Further
we find an exponential truncation of the scale-free behavior for $k > k_{\times}
\approx 2\times 10^5$, in contrast with the plateau reported in other
studies \cite{BarabasiRMP,Bianconi01}.  For the second database, we also find
a power-law regime with the same exponent, but the exponential truncation
appears to be absent, suggesting that the truncation is not due to the finite
size of the databases.

To explain these empirical results, we hypothesize that the authors of new
web-pages filter some of the information regarding existing web-pages,
that is, the new nodes make linking decisions under information-filtering
conditions.  To investigate this process, we consider network growth models
in which new nodes process information from only a fraction of existing
nodes which one may view as matching the ``interests'' of the new nodes.
If the fraction $f$ of ``interesting'' nodes in the network
is much less than one, then the attachment of new links is a
random process, so the generated network will be a random graph with an
exponentially-decaying in-degree distribution. In contrast, if $f \approx 1$,
then preferential attachment is recovered and the in-degree distribution is
scale-free.

We first define the network growth rule: At time $t=0$,
one creates $n_o$ nodes with $n_o-1$ links each. At each time step, one
adds to the network a new node with $n_o-1$ outgoing links.  These $n_o$
links can connect to a randomly selected subset ${\cal C}$ containing
$n(t) = (t+n_o)f$ nodes.  The links to the nodes in the subset are
selected according to the preferential attachment rule, i.e., the
probability that node $i$ belonging to ${\cal C}$ is selected is
proportional to the number of incoming links $k(i)$ to it
\begin{equation}
p(i,t) \equiv \frac{k(i)}{\sum_{j\epsilon {\cal C}} k(j)} \,.
\end{equation}

\begin{figure}[t]
\centering
\includegraphics[width=.4\textwidth]{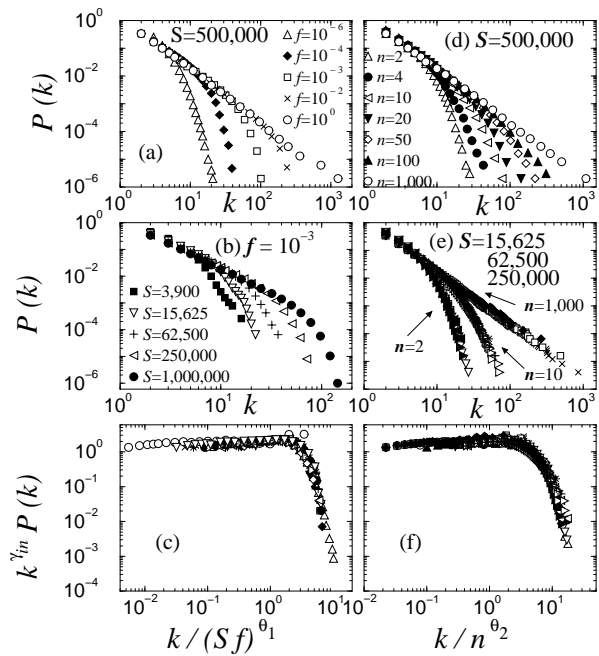}
\vspace{-0.4cm}
\caption{ \label{fig2} In-degree cumulative probability distributions $P(k)$ under 
information filtering. {\it Constant $f$ case:} 
(a) Results for $S=5\times 10^5$ and different values of $f$.
(b) Results for $f=10^{-3}$ and different values of $S$.
(a) and (b) show that $k_{\times}$ decreases with $f$
and increases with $S$.
(c) Data collapse of the numerical results according to
Eq.~(\protect\ref{e.scaling-function-f}) with $\gamma_{in}=1.97\pm 0.05$ and
$\theta_1 = 0.45 \pm 0.04$.
{\it Constant $n$ case:}
(d) Results for $S=5\times 10^5$ and different values of $n$ showing the decrease
in the cut-off degree $k_{\times}$ with decreasing $n$.
(e) Results for $n =2, 10$ and $1,000$ for different values of $S$ 
showing that $P(k)$ does not depend on $S$.
(f) Data collapse according to
Eq.~(\protect\ref{e.scaling-function-n}) with $\gamma_{in} = 2.00 \pm 0.03$ and
$\theta_2 = 0.65 \pm 0.04$.
}
\end{figure}

In Fig.~\ref{fig2}(a), we show our numerical results for the in-degree
cumulative distributions for networks with $S=5 \times 10^5$ nodes and $n_o = 1$,
for a sequence of $f$ values.  For $f=1$, we reproduce the results reported
for the scale-free model~\cite{Barabasi99}---i.e. we observe an
in-degree distribution that decays as a power law with an exponent
$\gamma_{in} \approx 2$.  For $f < 10^{-2}$, we observe a crossover at $k =
k_\times$ from power-law behavior to exponential behavior.

To further investigate the effect of changes in $f$ on the cut-off degree
$k_\times$, we plot in Fig.~\ref{fig2}(b) the in-degree distributions
for different network sizes $S$ and a fixed value of $f$. We find
that $k_\times$ increases as a power law with 
$S$. All of our numerical results can be expressed compactly by the scaling
form
\begin{equation}
P(k,f,S)\propto k^{-\gamma_{in}} {\cal F}_1 
\left(\frac{k}{k_{\times}}\right)
\label{e.scaling-function-f}
\end{equation}
with $k_{\times} \sim (Sf)^{\theta_1}$. We find $\gamma_{in} = 1.97 \pm 0.05$,
$\theta_1 = 0.45 \pm 0.04$ and ${\cal F}_1(x) \sim {\rm const}.$ for $x \ll
1$, ${\cal F}_1(x) \sim e^{-x} $ for $x \gg 1$. 
As a test of the scaling form Eq.~(\ref{e.scaling-function-f}), we plot in
Fig.~\ref{fig1}(c) the scaled cumulative distribution 
versus the scaled in-degree.  The figure confirms our
scaling Ansatz, since all data ``collapse'' onto a single curve, the scaling
function ${\cal F}_1(x)$.


We consider next a situation in which new nodes are not processing
information from a constant fraction $f$ of nodes but from a constant
number $n$ of nodes.  That is, as the network grows, the new nodes are
able to process information about a smaller fraction of
existing nodes. This model may be more plausible for networks that
have grown to a very large size, since the
fraction $f$ of all nodes represents a very large number.  In
the case of the scientific literature, this effect leads to the
fragmentation of a scientific field as it grows \cite{vanRaan00}.  

For the constant $n$ case, the fraction of known nodes at time $t$ is
$f(t) = n/(t+n_o)$, implying that as the networks grows there are two
antagonistic trends affecting $k_{\times}$.  The first is a tendency
to increase due to the growing size of the network, and the second is
a tendency to decrease due to the decreasing value of $f$.  Hence, one
may hypothesize that there will be a characteristic network size $S_c$
above which $k_{\times}$ will no longer depend on $S$.

\begin{figure}[t]
\includegraphics[width=.4\textwidth]{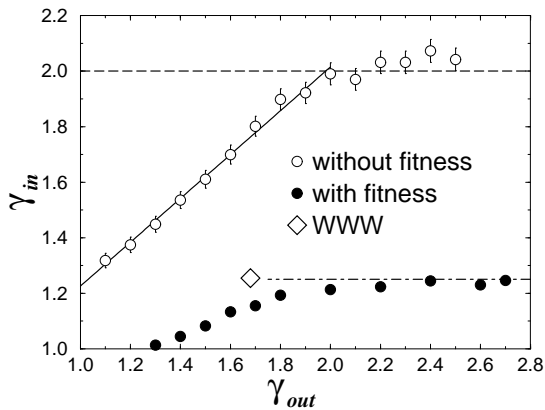}
\vspace{-.5cm}
\caption{ \label{fig4}
Dependence of the in-degree distribution exponent $\gamma_{in}$ on the
out-degree distribution exponent $\gamma_{out}$. We show results for models
{\it (i)} without fitness ($\eta(i)={\rm const.}$)
and {\it (ii)} with fitness ($\eta(i)$ uniformly distributed).  For the former case,
$\gamma_{in}$ increases initially approximately linearly with $\gamma_{out}$,
and then saturates at $\gamma_{in} \approx 2$ for $\gamma_{out} > 2$.  This
saturation of $\gamma_{in}$ is to be expected as $\gamma_{in}=2$ 
for the case of a peaked distribution of $n_o$.  For the latter
case, $\gamma_{in}$ increases approximately linearly with 
$\gamma_{out}$ initially, and then saturates at
$\gamma_{in} \approx 1.25$ for $\gamma_{out} > 1.9$.  This saturation 
is to be expected as $\gamma_{in}=1.255$ for
the case of a peaked distribution of $n_o$ \protect\cite{Bianconi01}.
}
\end{figure}

We now test these arguments with numerical simulations.  In
Fig.~\ref{fig2}(d)-(e), we show our results for growing networks for
which new nodes process information only from $n$ randomly selected
existing nodes.  We find, in agreement with our scaling arguments,
that for $S\gg S_c$ the in-degree distribution obeys the scaling
relation
\begin{equation}
P(k,n,S)\propto k^{-\gamma_{in}} {\cal F}_2 \left(\frac{k}{k_{\times}}\right) \,,
\label{e.scaling-function-n}
\end{equation}
with $k_{\times} \sim n^{\theta_2}$, $\gamma_{in} =2.00 \pm 0.03$, $\theta_2 =
0.65 \pm 0.04$, and where the scaling function
${\cal F}_2(x)$ has the same limiting behavior as ${\cal
F}_1(x)$. To test the scaling form Eq.~(\ref{e.scaling-function-n}), we plot 
in Fig.~\ref{fig2}(f) the scaled cumulative distribution 
versus the scaled in-degree.  This confirms our
scaling Ansatz since the data collapse onto a single curve, the scaling
function ${\cal F}_2(x)$.


Comparison of the two scaling relations Eq.~(\ref{e.scaling-function-f}) and
Eq.~(\ref{e.scaling-function-n}) reveals an unexpected result.  By replacing $Sf$
by $n$ in (\ref{e.scaling-function-f}) one would naively expect to obtain
(\ref{e.scaling-function-n}) with $\theta_1 = \theta_2$ and ${\cal F}_1(x) =
{\cal F}_2(x)$.  Surprisingly, we find that $\theta_1$ is significantly
different from $\theta_2$ and that ${\cal F}_1(x)$ is significantly different
from ${\cal F}_2(x)$.
In order to understand this result, consider two growing networks
that have reached size $S$.  For the first, new nodes process information
from a fraction $f$ of existing nodes, while, for the second, new nodes
process information from $n = fS$ existing nodes.  At a time $t$, 
prior to the network having reached its final size $S$,
there are
$t+n_o < S$ sites, and the preferential attachment is acting for the first
network on a number of nodes $(t+n_o)f < Sf = n$. The preferential attachment
mechanism can operate effectively only when it acts on a number of nodes
comparable to $S$, so the fact that for the first network new nodes have
always processed information from {\it fewer\/} existing nodes suggests the
first network will not develop nodes with as large a degree as the second
network. Thus, we expect that (i) the two resulting networks have different
in-degree distributions, and (ii) the in-degree distribution for $f$ fixed 
has a sharper truncation and a smaller cut-off than for $n$ fixed, which is
indeed what we find.

%
Our numerical results are in qualitative agreement with empirical
data. However, the value of the power law exponent $\gamma_{in} \approx 1.25$
found for the WWW is significantly smaller than the value $\gamma_{in} =2$
predicted by the model.  This fact prompts the question of the effect of the
cost of information filtering on models generating an in-degree
distribution closer to the empirical results.  To answer this question, we
investigate two possible explanations for the observed value $\gamma_{in}
\approx 1.25$.  

{\em (i) Effect of out-degree distribution on $\gamma_{in}$.}
The scale-free model\cite{Barabasi99} is missing an important ingredient: a heterogeneous
distribution of number of outgoing links. Indeed, the out-degree distribution
considered so far is restricted to a single value $m = n_o-1$
, i.e. $p_{out}(m) = \delta_{m,n_o-1}$, while
for the empirical data of the WWW it decays as a power law 
of the form $p_{out}(m)\sim m^{-\gamma_{out}}$
with $\gamma_{out} = 1.68 \pm 0.05$.
We show in Fig.~\ref{fig4} the computed value of 
the exponent $\gamma_{in}$ of the in-degree
distribution as a function of $\gamma_{out}$~\cite{gammanote}.  
We find that $\gamma_{in}$ increases approximately
linearly with increasing values of the exponent $\gamma_{out}$ until it reaches
the limiting value $\gamma_{in} = 2$.  For $\gamma_{out} \approx 1.7$, which is
the empirically-observed value for the WWW, we find $\gamma_{in} \approx 1.8$,
which does not agree with the empirical value of 1.25, so 
the power-law decaying out-degree distribution alone cannot explain the
results obtained for the WWW.

{\em (ii) Effect of fitness on $\gamma_{in}$.}
The preferential attachment mechanism is modified by 
a ``fitness'' factor \cite{Bianconi01}: Nodes
have different fitness, and fitter nodes are more likely to receive
incoming links than less fit nodes with the same value of $k$.
Uniformly-distributed fitness is known to 
lead to a smaller exponent $\gamma_{in} = 1.255$
\cite{Bianconi01}, which is quite close to the value measured for the
WWW. Hence, we assign to each node a fitness $\eta(i)$ \cite{Bianconi01},
reflecting the fact that for equal values of $k$ some nodes are
more ``attractive'' than others\cite{footnote}. The probability that a
new node will link to node $i$ is
\begin{equation}
p(i,t)\equiv \frac{\eta(i)k(i)}{\sum_{j=1}^{t+n_o}\eta(j)k(j)} \,.
\label{e-fitness}
\end{equation}
We consider here the case in which $\eta(i)$ is a uniformly distributed
random variable~\cite{fitnessnote}.  Figure~\ref{fig4}  
shows that the in-degree distribution
decays as a power law with values of $\gamma_{in} < 1.25$. For $\gamma_{out}
> 1.9$, the exponent approaches the limiting value $\gamma_{in} \approx
1.25$.  Interestingly, for $\gamma_{out} \approx 1.7$, the empirical value
for the WWW, we find $\gamma_{in} \approx 1.2$, in agreement with the
empirical value $\gamma_{in} \approx 1.25$.

Our results for the model with fitness show that
information filtering and node fitness are both
necessary in order to approximate the empirical results.
An open question is which type of filtering is more
appropriate for the WWW, constant $f$ or constant $n$?  To answer this
question one would need WWW data for a different sample size, which are not
available to us at present.  However, due to the sheer size of the WWW, it
seems plausible that constant $n$ would be the more appropriate case.

Our key finding is that limited information-processing
capabilities have a significant and quantifiable 
effect on the large-scale structure of
growing networks. We find that
information filtering leads to an exponential truncation of the in-degree
distribution for networks growing under conditions of preferential
attachment.  Surprisingly, we find simple scaling relations that predict the
in-degree distribution in terms of {\em (i)} the information-processing
capabilities available to the nodes, and {\em (ii)} the size of the network.

We also quantify the effect of a heterogeneous out-degree distribution on the
in-degree distribution of networks growing under conditions of preferential
attachment.  We find that for a power law decaying out-degree distribution
with exponents $\gamma_{out} < 2$, the exponent $\gamma_{in}$ characterizing
the tail of the in-degree distribution will take values smaller than those
predicted by theoretical calculations \cite{BarabasiRMP,MendesRev}.

The exponential truncation we find may have dramatic effects on the dynamics
of the system, especially for processes where the nodes with the largest
degree have important roles. This is the case, for example, for virus
spreading \cite{Cohen94}, where for networks with exponentially-truncated
in-degree distributions there is a non-zero threshold for the appearance of
an epidemic.  In contrast, scale-free networks are prone to the spreading and
the persistence of infections no matter how small the spreading rate. Our
finding of a mechanism leading to an exponential truncation even for systems
where before none was expected \cite{Amaral00} indicates that the most
connected nodes will have a smaller degree than predicted for scale-free
networks leading, possibly, to different dynamics, e.g., for the initiation and
spread of epidemics.

In the context of network growth, the impossibility of knowing the degrees of
all the nodes comprising the network due to the filtering process
---and hence the inability to make the
optimal, rational, choice---is not altogether unlike the ``bounded
rationality'' concept of Simon \cite{Simon97}.  
Remarkably, it appears that for the description of WWW growth, the
preferential attachment mechanism, originally proposed by Simon \cite{Simon},
must be modified along the lines of another concept also introduced by
him---bounded rationality \cite{Simon97}.

We thank R.~Albert, P.~Ball, A.~-L.~Barab\'asi, M.~Buchanan,
J.~Camacho, and R.~Guimer\`a for stimulating discussions
and helpful suggestions.  We are especially grateful to R.~Kumar for sharing
his data. We thank NIH/NCRR (P41 RR13622) and NSF for support.


%

\begin{references}
\bibitem{gammanote} We consider a modification to the network growth rule
described earlier in the paper: at each time step $t$ the new node establishes 
$m$ new links where $m$ is drawn from a power law distribution with 
exponent $\gamma_{out}$.
\bibitem{fitnessnote} 
It is known~\cite{Bianconi01} that 
for an exponential or fat-tailed distribution of fitness,
the structure of the network becomes much more complex; 
in particular the in-degree distribution is no longer a power law.
Hence we do not consider in this manuscript 
other shapes of the fitness distribution.
\end{references}
\end{document}